# Quantum oscillations revealing topological band in kagome metal ScV$_6$Sn$_6$

Changjiang Yi[1], Xiaolong Feng[1], Ning Mao[1], Premakumar Yanda[1], Subhajit Roychowdhury[1], Yang Zhang[2,3], Claudia Felser[1], and Chandra Shekhar*[,1]

[1]Max Planck Institute for Chemical Physics of Solids, 01187 Dresden, Germany
[2]Department of Physics and Astronomy, University of Tennessee, Knoxville, TN 37996, USA
[3]Min H. Kao Department of Electrical Engineering and Computer Science, University of Tennessee, Knoxville, TN 37996, USA

Compounds with kagome lattice structure are known to exhibit Dirac cones, flat bands, and van Hove singularities, which host numerous versatile quantum phenomena. Inspired by these intriguing properties, we investigate the temperature and magnetic field dependent electrical transports along with the theoretical calculations of ScV$_6$Sn$_6$, a non-magnetic charge density wave (CDW) compound. At low temperatures, the compound exhibits Shubnikov–de Haas quantum oscillations, which help to design the Fermi surface (FS) topology. This analysis reveals the existence of several small FSs in the Brillouin zone, combined with a large FS. Among them, the FS possessing Dirac band is a non-trivial and generates a non-zero Berry phase. In addition, the compound also shows the anomalous Hall-like behaviour up to the CDW with the CDW phase, ScV$_6$Sn$_6$ presents a unique material example of the versatile HfFe$_6$Ge$_6$ family and provides various promising opportunities to explore the series further.

The kagome lattice gives rise to unavoidable exotic topological electronic states namely Dirac point [1, 2], Van Hove singularity [3, 4] and flat band [1, 2]. These features have been extensively studied by different spectroscopy experiments in various kagome compounds. Depending on band filling and interactions, they accommodate a number of unconventional quantum phases, topological band structure [5–7], Chern insulator [8], unconventional superconductivity [9,10] and charge density wave (CDW) [11–14]. The profound impacts of the kagome lattice have recently been highlighted by the discovery of CDW below 90 K in nonmagnetic $A$V$_3$Sb$_5$ and antiferromagnetic hexagonal-FeGe [11]. The charge orders in these systems are distinct. In $A$V$_3$Sb$_5$ [14–18], it is chiral, whereas in FeGe [11,19] magnetism is mediated. Similar to $A$V$_3$Sb$_5$ and FeGe, the CDW transition temperature (at 92 K) was recently discovered in the ScV$_6$Sn$_6$ compound [20]. ScV$_6$Sn$_6$ is the only known compound from the vast hexagonal HfFe$_6$Ge$_6$ family to exhibit a CDW phase transition, which is a first-order-like transition with propagation vector (1/3, 1/3, 1/3) revealed by early X-ray and neutron experiments [20]. The strongly coupled out-of-plane lattice dynamics suggests an unconventional nature of the CDW phase, which is different from the in-plane lattice dynamics in $A$V$_3$Sb$_5$ [12]. Furthermore, CDW of ScV$_6$Sn$_6$ is easy to tune [21] and exhibits various microscopic features such as critical role of phonons [22–25], large spin Berry curvature [26], partial bandgap opening [26–28], hidden magnetism [29]. As the first such member of the versatile kagome family of HfFe$_6$Ge$_6$-type compounds, it is worthwhile to study the temperature and magnetic field dependent electrical transport properties of high-quality single crystals of ScV$_6$Sn$_6$ in order to obtain information about the Fermi surface (FS) and lattice dynamics, reporting in this letter.

ScV$_6$Sn$_6$ crystallizes at room temperature in a hexagonal centrosymmetric structure with a *P*6/*mmm* space group. The V atoms form a kagome lattice in the *ab*-plane (Fig. 1(a) left), while the Sn atoms sit above and below, separating the lattice [20]. High quality hexagonal single crystals (Fig. 1(b), inset) were grown by the flux method, see supplemental information (SI) [30].

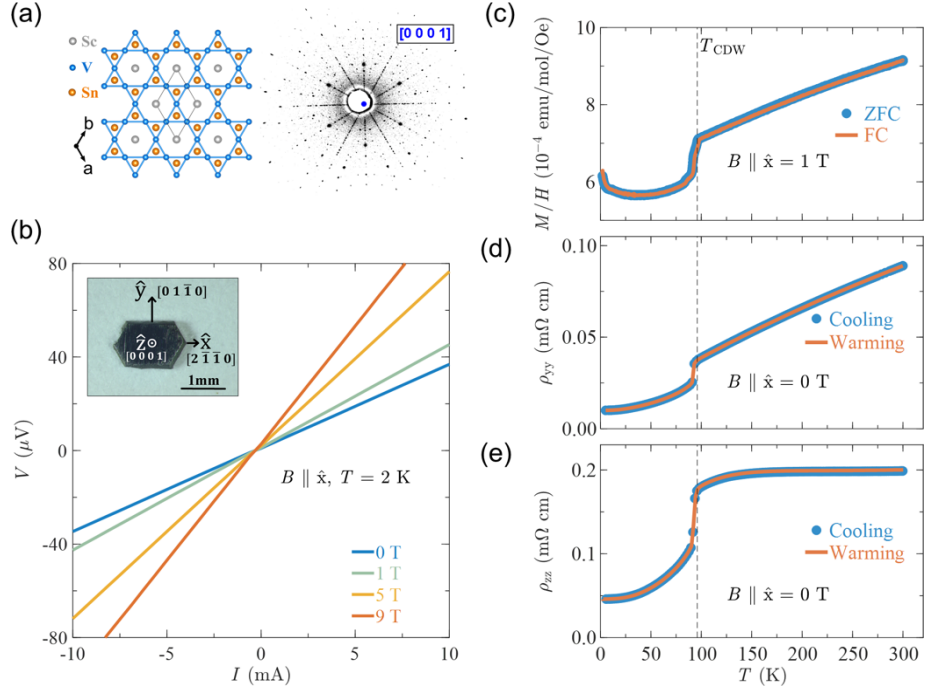

FIG. 1. Kagome lattice, Laue pattern, I-V characteristic, CDW phase transition of ScV$_6$Sn$_6$. (a) Kagome lattice of ScV$_6$Sn$_6$ viewed along the *c*-axis (left) and recorded single crystal Laue patterns along the *c*-axis. (b) I-V characteristics at various fields at 2K. The inset is an optical image of the crystal with Cartesian coordinate. (c) Magnetic susceptibility along $\hat{x}$. Resistivity in zero-field (d) along $\hat{y}$ and (c) along $\hat{z}$, where the sudden jumps show CDW phase transition.

To facilitate further measurements, the crystal orientations were marked in Cartesian coordinates $\hat{x}$, $\hat{y}$, $\hat{z}$, which correspond to $(2\bar{1}\bar{1}0)$, $(01\bar{1}0)$, $(0001)$ crystallographic directions in the hexagonal crystal structure, respectively. Laue X-ray diffraction patterns were recorded along $\hat{z}$ and show a clear six-fold symmetry (Fig. 1(a), right). The patterns fit well with the lattice parameters $a = b = 5.4733(5)$ Å and $c = 9.1724(8)$ Å derived from the fitting of power X-ray patterns (Fig. S1) [30,31]. The measured chemical compositions from the energy dispersive X-ray spectrum are close to the stoichiometric atomic ratio of Sc:V:Sn = 1:5.97:6.07. We further cut the crystals in defined orientations of Hall-bar geometry and fabricated four-probe Pt-wire contacts with highly conductive Ag paint. The *I-V* characteristics at 2 K in different fields are linear (Fig. 1(b)), confirming the high quality of the ohmic contacts. The appearance of CDW in this compound is a key property, and we found it in the temperature-dependent magnetization and the electrical resistivity (Figs. 1(c-e)). A sharp transition appears at 92 K, depicting the CDW transition temperature similar to the previous reports [19, 20]. The measured magnetic susceptibility in Fig. 1(c) describes a weak Pauli paramagnetic-like behavior, which is different from its structural sister compound $YV_6Sn_6$ [32]. The zero-field electrical resistivity shows a metallic behavior with temperature, and both $\rho_{yy}$ and $\rho_{zz}$ drop suddenly at the CDW phase transition due to the large softening of the acoustic phonon modes (Figs. 1(d, e)) [23,24]. The typical values of $\rho_{yy}$ ($\rho_{zz}$) at 2 K are found to be $1 \times 10^{-5}$ ($4.58 \times 10^{-5}$) $\Omega$ cm and the resulting residual resistivity ratio ($RRR = \rho_{300K}/\rho_{2K}$) is 8.9(4.3). Noticeably, the $\rho_{zz}$ changes significantly than the $\rho_{yy}$ at the CDW phase transition.

After the basic characterization of the crystals, we now focused on the magnetic field, *B*-dependent transverse resistivity, which were measured at different temperatures and angles in the field of ± 9 T. The angular dependence of the field is defined as $\theta$ (= $\hat{x} \to \hat{z}$) and $\phi$ (= $\hat{x} \to \hat{y}$), where 0° = $B \parallel \hat{x}$, 90° = $B \parallel \hat{y}$ or $\hat{z}$. First, we describe the behavior of $\rho_{zz}$ when $B \parallel \hat{x}$ ($\theta$ and $\phi$ = 0), and the measured data are shown in Fig. S6(a) [30]. At low temperatures, $\rho_{zz}$ exhibits clear Shubnikov de-Haas (SdH) quantum oscillations which are a striking feature that helps to probe the low-energy bands. At 2 K, the quantum oscillation starts from the field around 1.5 T, as seen in Fig. 2(a) for the $d\rho_{zz}/dB$ plot and it is easily visible up to 12 K (Fig. S6(b) [30]. The magnetoresistance MR $(= \rho_{zz}(B)/\rho_{zz}(0) - 1)$ is estimated, and the highest value is found to be 170 % at the lowest temperature of 2 K (Fig. S6(a)) [30]. Unlike the other typical semimetals, e.g., NbP, the present compound $ScV_6Sn_6$ does not show an extremely large MR [33]. However, the SdH oscillation indicates a low effective mass of the electron charge carrier in the CDW phase [23,24]. To obtain the amplitude of the SdH oscillations, a smooth polynomial background was subtracted from the measured $\rho_{zz}$. The resultant $\Delta\rho_{zz}$ at several temperatures is plotted as a function of $1/B$, as shown in Fig. 2(b). As expected, the oscillations are periodic in $1/B$ and they arise from quantization of the energy level further, forming Landau levels (LLs) [34]. The analysis of the temperature and field-dependent periodic oscillations provides insight into the fermiology of the compound and the associated physical characteristics of the charge carriers.

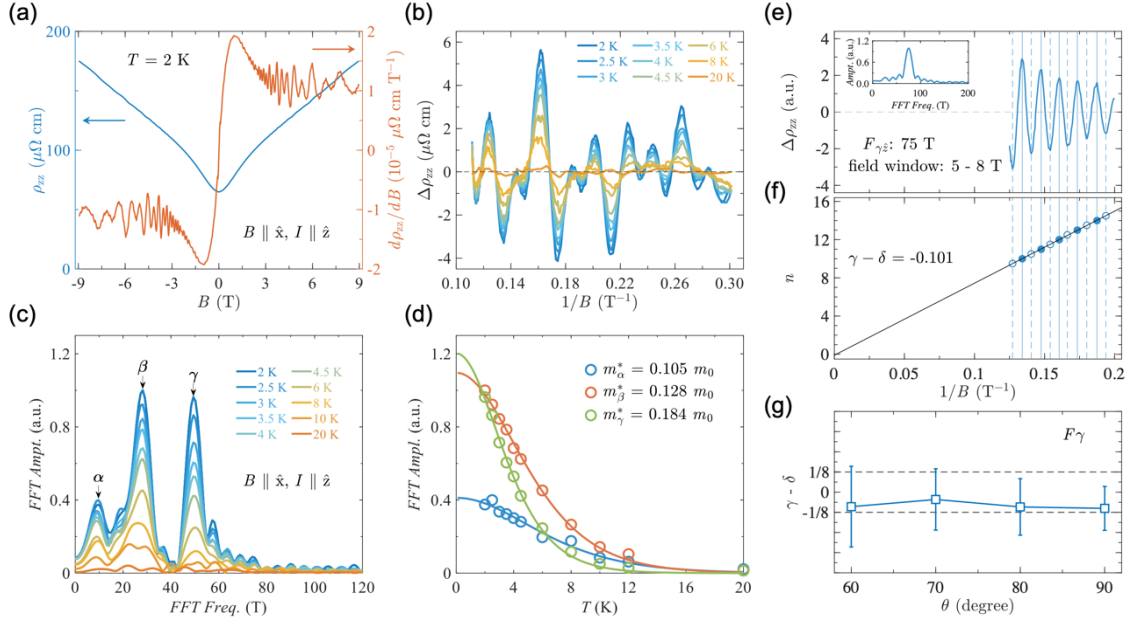

FIG. 2. Shubnikov–de Haas (SdH) oscillations and Berry phase. (a) Longitudinal $\rho_{zz}$ at 2 K and its first derivative with field showing quantum oscillations. (b) Background subtracted SdH oscillations amplitude at several temperatures. (c) Corresponding FFT amplitude exhibiting $\alpha$, $\beta$ and $\gamma$ frequencies. (d) FFT amplitude fit estimating effective mass. (e) SdH oscillations when $B \parallel \hat{z}$ and their FFT spectrum (inset) showing only the $\gamma$ frequency in the field window of 5-8 T. (f) Intercept of the Landau fan diagram revealing non-zero Berry phase. (g) Angular dependent of the Berry phase.

In the fast Fourier transform (FFT) of the SdH oscillations presented in Fig. 2(c), we detected the existence of three distinct frequencies: $F_\alpha = 10$ T, $F_\beta = 28$ T and $F_\gamma = 50$ T. These frequencies indicate the presence of small FSs. Compared to $AV_3Sb_5$ (A = K, Rb, Cs), the splitting of FS in $ScV_6Sn_6$ is simpler. At least 9 FSs have been detected in $CsV_3Sb_5$, with their belonging frequency ranging from 10 T to 2000 T [35–38]. Lifshitz–Kosevich (L-K) theory is used extensively to analyze the physical properties of FSs [34], and it fully explains the SdH oscillations. The effective mass, Dingle temperature and phase factor are evaluated using L-K equation, where $\Delta\rho \propto R_T R_D \cos\left[2\pi\left(\frac{F}{B} + \gamma - \delta\right)\right]$. Specifically, $R_T = 14.69 m^* T/B \times \sinh(14.69 m^* T/B)$ and $R_D = \exp(-14.69 m^* T_D/B)$ are the cyclotron effective mass term and Dingle term, respectively; in which $m^*$ ($m_0$) is the effective (bare) mass of the electron, $B$ is the average field used in the FFT, $T_D$ is the Dingle temperature. The phase factor $\gamma - \delta$, which is directly related to the Berry phase within $\delta$, equals to 0 for the 2D system and $\pm\frac{1}{8}$ for the 3D system ( $\pm$ corresponds to the contribution from the minimal/maximal cross section) [36,39,40]. Among the various parameters of FS that can be estimated using L-K formula, the estimation of the Berry phase is very promising for proving the non-trivial topology of the band. Accordingly, we designed the Landau fan diagram by assigning LLs to the oscillatory extrema. The integer $n$-th LLs are assigned to maxima when $\rho_{xx} \gg \rho_{yx}$, while they are assigned to minima when $\rho_{xx} \ll \rho_{yx}$ [36,39,41]. For ScV$_6$Sn$_6$, where $\rho_{zz} \sim 10 \rho_{yz}$ which satisfies the condition of $\rho_{xx} \gg \rho_{yx}$, the integer $n$-th are assigned to the maxima and the half integer $(n + \frac{1}{2})$-th are assigned minimum of $\Delta\rho_{zz}$, as shown in Fig. 2(e, f)). The Landau indices $n$ and $1/B$ satisfy the Lifshitz-Onsager relationship, which is described by the equation $n = \frac{F}{B} + \gamma - \delta$. The slope of the plot between $n$ and $1/B$ provides the oscillatory frequency, while the intercept offers the information on the Berry phase. To attend an accurate estimation of Berry phase of $F_\gamma$, we selected a field window between 5 to 8 T to exclude the weak $F_\alpha$ oscillatory component (Fig. 2(e) inset). From Fig. 2(f), the intercept of the Landau fan diagram for $F_\gamma$ in terms of the phase factor $\gamma - \delta$ is estimated to be -0.101, indicating a nontrivial Berry phase [42,43]. Similarly, we obtained the Berry phase further for the angular dependence of the field from $\hat{x}$ to $\hat{z}$ without interference from the other exiting frequencies, and the values are given in Fig. 2(g). The figure indicates the Berry phase factor $\gamma - \delta$ falls within the range of $0 \pm \frac{1}{8}$ for $\theta \geq 60°$. While it is possible to estimate the Berry phase factor $\theta > 20°$, since the $F_\beta$ is absent, but the SdH amplitude declines rapidly. At all other angles of $\theta$ and $\phi$, the $F_\beta$ is consistently present as can be seen in Fig. 4(b), which intervenes the amplitude in terms of the beating patterns [44,45], as can be seen in Figs. S9(c) and (d) [30]. One method to separate the amplitude corresponding each frequency is to use the band-passing filter as shown in Figs. S6(c) and (d) [30,46]), but may not be very reliable. We also tried to estimate to Berry phase related to $F_\alpha$ and $F_\beta$. As we can see, $F_\alpha$ shows very weak oscillatory amplitude with limited cycle up to 9 T. One the other hand, the presence of the beating patterns harms the amplitude of the $F_\beta$, which only appears for $\theta \leq 20°$ and after that the $F_\beta$ vanishes. These together make difficult to carry out the Berry phase analysis for $F_\alpha$ and $F_\beta$.

To determine the effective mass for the carriers corresponding to each FS, the temperature dependence of the amplitude of each peak in the FFT is plotted, and then it is fitted with $R_T$ term (Fig. 2(d)) from the L-K formula. From these fittings, we find the values of $m^*_{\alpha,\beta,\gamma}$ = 0.105 $m_0$, 0.128 $m_0$, 0.184 $m_0$. These masses are very low (even lower than $CsV_3Sb_5$ [35,36]). Furthermore, the Dingle temperature $T_D$ can be estimated from the semi-log plot described by $\ln(A/R_T) \propto -14.69 m^* T_D/B$ (Fig. S8) [30], where $A$ corresponds to the amplitude of the SdH oscillations. We found $T_D$ to be 2.5 K for $\gamma$–pocket and the corresponding quantum scattering time $\tau_q$ is $\sim 5 \times 10^{-13}$ s. For the $F_{\alpha,\beta,\gamma}$ = 10, 28, and 50 T, the equivalent periodicities $1/(F_{\alpha,\beta,\gamma})$ = 0.1, 0.036, and 0.02 T$^{-1}$, which are further related to the cross-sectional area $A_F$ of FS by the Onsager relation $F = (\Phi_0/2\pi^2)A_F$, where the value of magnetic flux quantum $\Phi_0$ = 2.068 × 10$^{-15}$ Wb. The corresponding area $A_{F\alpha,\beta,\gamma}$ = 0.001, 0.0027, and 0.0048 Å$^{-2}$. Assuming a circular cross-section of these FSs for simplicity, the very small Fermi vectors $k_{F\alpha,\beta,\gamma}$ = 0.017 Å$^{-1}$, 0.029 Å$^{-1}$, 0.039 Å$^{-1}$ are found; and their corresponding Fermi velocities $v_{F\alpha,\beta,\gamma}$ are $1.92 \times 10^5$ m s$^{-1}$, $2.64 \times 10^5$ m s$^{-1}$, $2.45 \times 10^5$ m s$^{-1}$. Additional relevant experimental parameters of the FSs are summarized in Table S1 [30].

To capture the topology in electronic band structure of ScV$_6$Sn$_6$, we performed detailed band structure calculations by using density functional theory (DFT). Due to the presence of CDW transition, we transformed the band structure of low temperature (LT) phase into the high-temperature (HT) phase, and the LT phase (CDW phase) has been characterized at a temperature of 50 K and the lattice parameters are taken from the previous report [20]. From Fig. S11 [30], the unfolded band structures of LT and HT phases are in a good agreement with the angle-resolved photoemission spectrum (ARPES) experiment [27]. Among the two phases, the major change in the electronic band structure appears only along the *A-L-H* direction (where the band opens a gap because of the CDW transition [27]), while the Van Hove singularity bands at the *M* point and the Dirac bands at the *K* point do not change. From a closer view, a fascinating feature depicting in Fig. 3(a) is the band folding phenomenon, whereby the original *K* point at coordinates (1/3, 1/3, 0) transmutes into a folded *K*1 point situated at (1/3, 1/3, -2/3). This figure unambiguously reveals that the *K*1 point intrinsically possesses a substantial band gap near the Fermi energy ($E_F$). Furthermore, both the conduction and valence bands are completely split at this point, implying the absence of topologically protected Dirac points. Utilizing the transformation matrix, we ascertain that the original *M* point at (1/2, 0, 0) and the *Γ* point at (0, 0, 0) have been mapped to the new *M*1 point at (1/2, 0, -1/2) and *Γ* point at (0, 0, 0), respectively. As a result, we discovered that the genuine Dirac point (labeled as DP in Fig. 3b) has been mapped to the position of (0.341666, 0.316666, -0.658327). From this magnified view of the band structure near this point, its Dirac nature is clearly confirmed. Similar to the Dirac points in graphene systems, this point develops a band gap when considering spin-orbit coupling as shown in Fig. 3(c), and gives rise to non-trivial topological properties. Therefore, we believe that the non-trivial π Berry phase extracted in the SdH oscillation actually comes from the electron pockets formed in the energy bands where the DP is located.

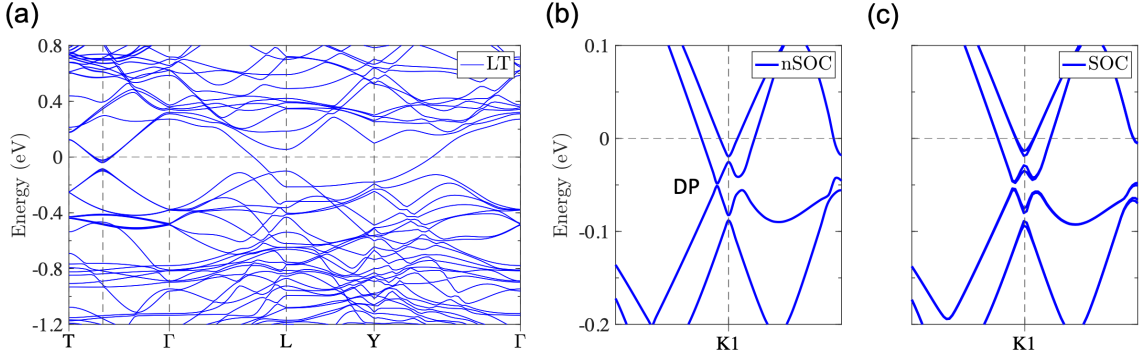

FIG. 3. Band structure in charge density wave (CDW) phase. (a) Band structure of $ScV_6Sn_6$ CDW phase with the presence of spin orbital coupling (SOC). The structural parameters are the same with the report [20] at 50 K. The magnified view of the band structure around $K1$ locating the Dirac point (DP) (b) without the SOC and (c) with the SOC.

To know more information about the topography of the FS, we measured the angular $\theta\,(=\hat{x} \rightarrow \hat{z})$ and $\phi\,(=\hat{x} \rightarrow \hat{y})$ dependent oscillations. The rotating angles of field with the crystal axes are sketched in the insets of Fig. 4(a), where $0° = B \parallel \hat{x}$, $90° = B \parallel \hat{y}$ or $\hat{z}$. The angular dependent of FFT is plotted in Fig. 4(a). As we mentioned earlier, the three frequencies namely α, β and γ appear when $B \parallel \hat{x}$. Depending upon the shape of FS, the movement of frequency depends on the rotating angle. From Fig. 4(a), we carefully tracked the position of all visible frequencies with the rotating angles and demonstrated in Fig. 4(b). This indicates that when the field $B$ moves either from the $\hat{x} \rightarrow \hat{y}$ or $\hat{x} \rightarrow \hat{z}$, the frequencies α, β and γ change slightly. Noticeably, when the $\theta > 20°$, the β disappears. This helps further to improve the accuracy of Berry phase analysis related to γ. In order to allocate the experimentally observed frequencies, we theoretically calculated 3D FSs and their corresponding frequency from the unfolded band structure of the LT phase. As we can see in Fig. 4(b), all observed frequencies are accurately replicated by the calculated 3D FSs, except for higher frequency (green line). In addition, the calculated $F_\beta$ is present at all angles, but it disappears $\theta > 20°$ in the experiment, which needs a further investigation. Fig. 4(c) shows 3D FSs within the BZ, in which a large electron pocket (blue color) around $\Gamma$ covers the whole BZ together with a small banana-shaped FS (red color) along the $M$-$L$ direction, a small apple-shaped FS (blue color) at $K$, and a larger FS at $M$ (yellow color). Another spindle-shaped FS at $K$ which comes from the same band with the FS at $M$ and is covered by the apple-shaped pocket. After comparing with DFT results, we found that the $F_\alpha$ belongs to the banana-shaped FS (Fig. 4(e)). The $F_\beta$ belongs to the spindle-shaped FS (Fig. (4f)). The $F_\gamma$ is part of the apple-shaped FS that is contributed by the Dirac bands (Fig. 4(g)), exhibiting the non-trivial Berry phase. The calculated high frequency (green solid line in Fig. 4(b)) comes from the FS (Fig. 4(d)) at $M$ point, which is not detectable up to 9 T in the present measurements.

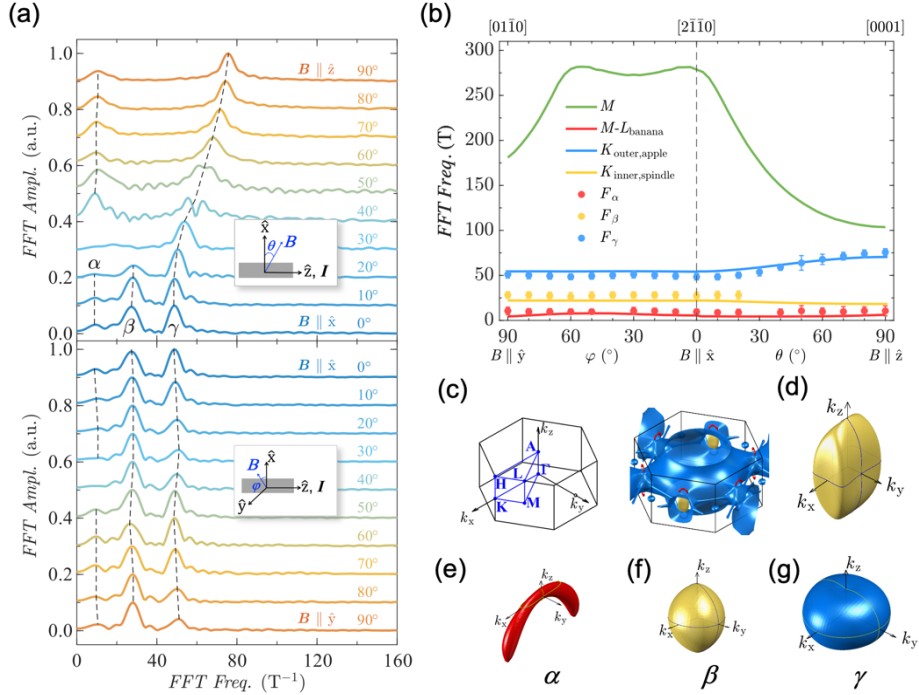

FIG. 4. SdH oscillations and fermiology. (a) SdH oscillation frequency at different rotating angles $\theta$ (= $\hat{x} \to \hat{z}$) and $\phi$ (= $\hat{x} \to \hat{y}$), where dotted lines represent their tracking with angle. The insets show a sketch of the field rotation. (b) Angular dependence of $F_{\alpha,\beta,\gamma}$, where $F_{\alpha,\beta}$ do not change while $F_\gamma$ shifts slightly. Solid lines are calculated frequencies corresponding. The error bar is taken from the half width of the half maximum value of the frequency peaks. (c) Brillouin zone and three-dimensional (3D) Fermi surfaces (FSs) of the bands at the Fermi energy. (d-g), Enlarged FSs, including the FS at $M$ point (d, yellow), the banana-shaped $\alpha$-pocket along $M$-$L$ (e, red), the spindle-shaped inner $\beta$-pocket at $K$ point (f, yellow), and the apple-shaped outer $\gamma$-pocket at $K$ point (g, blue).

After exploring the striking features of the FS, we now get the insight of classical Drude transport from the Hall resistivity $\rho_{zy}$ for current $I \parallel \hat{y}$ and $B \parallel \hat{x}$. From these data, the Hall constant $R_H$ ($=\rho_{zy}/B$) was calculated by considering the high field slope between 6–9 T at different temperatures (Fig. 5(a), inset), resulting in electron density $n$ ($= 1/eR_H$) and mobility $\mu$ ($= R_H/\rho_{yy}$) of $2.4 \times 10^{21}$ cm$^{-3}$ and 265 cm$^2$ V$^{-1}$ s$^{-1}$ at 2 K, respectively (Fig. S3(c)) [30]. The Drude scattering time, $\tau_{tr} = \mu m^*/e = 2.1 \times 10^{-14}$ s, where $m^*$ is 0.139 $m_0$ (the average of $m^*$) and $e$ is the elementary charge. The ratio of $\tau_c / \tau_{tr}$ (~ 25) is large, where $\tau_c$ and $\tau_{tr}$ are related to electron-electron and electron-phonon scattering processes, respectively. Besides the usual high and low angle scattering mechanisms, the large lowering value $\tau_{tr}$ could be related to the new phonon modes, arising from promoting the electron-phonon coupling crucially in the CDW phase [22-24].

The field dependent $\rho_{zy}$ shows a non-linear behavior, which is primarily thought to be the presence of multi-carriers in a nonmagnetic system such as ScV$_6$Sn$_6$, since the multiple bands are around the $E_F$. To clarify this apparent reason, we fitted both the Hall and longitudinal conductivities by using the standard two-band model (see Fig. S5 in SI) [30,47–50]. Both fits are seemingly good, but their fitted parameters are very different from each other. The fitting of the $\sigma_{yz}$ reflects both electron carriers, while the fitting of $\sigma_{yy}$ shows both electron and hole carriers. For the implementation of a reliable and accurate model, the parameters from the both fittings must be the same within the error bar as reported in the typical multiband systems [49,50]. Therefore, given the type of carriers, the concentrations and mobilities are highly inconsistent, indicating that the two-band model may not be applicable to the nonlinear Hall effect for ScV$_6$Sn$_6$. It is also worthy to note here that the Hall resistivity from a typical multi-band system is nonlinear close to zero field, while it is become linear at high field [33,51–53]. In contrast, the present compound ScV$_6$Sn$_6$ shows the linear Hall close to zero field and it bends at high field (> 2 T) like a typical soft ferromagnetic system [54,55].

By considering the anomalous behavior naively, we subtracted the linear part from the high field and the anomalous part of the Hall resistivity at different temperatures is shown in Fig. 5(a). Like in soft ferromagnet, an anomaly in the $\rho_{zy}$ (a sharp increase at a lower field followed by saturation with a further increase in the field) is observed, which is attributed to anomalous Hall like behavior. Such an anomaly usually appears as a hallmark in magnetic systems and it resembles the magnetization curve, wherein magnetic spins play the role (Fig. 5(d, e)). However, magnetization is linear with the field (see Fig. S2 in SI) [30] and does not follow the Hall resistivity in the present case. The as calculated value of anomalous Hall conductivity (AHC $\approx \frac{\rho_{zy}^A}{\rho_{zz}\rho_{yy}}$ since $\rho_{yy} \neq \rho_{zz}$) at 2 K is $3.2 \times 10^3$ $\Omega^{-1}$ cm$^{-1}$ which is smaller than $A$V$_3$Sb$_5$ systems [14,15,38]. The AHC remains almost constant up to 45º and starts to decrease with increasing angle further (Fig. 5(c)), while it sustains up to the CDW transition temperature. In contrast to the involvement of magnetism (Figs. 5(d), (e)), one of the possible explanations could be similar to the origin in $A$V$_3$Sb$_5$ (Fig. 5(f)), where the CDW forms a current loop and breaks the time reversal symmetry (TRS) [16,56,57], since ScV$_6$Sn$_6$ also breaks the TRS measured by the muon spectroscopy [29] and to speculate such loop current is a bit premature. In the present scenario, it is difficult to understand the observed Hall behavior of ScV$_6$Sn$_6$, and needs to do further study. Moreover, the behavior of $\rho_{yz}$ is closer to ZrTe$_5$, assuming the presence of a single Dirac band that splits into a pair of Weyl points in presence of the magnetic field [58]. However, the several bands are present in ScV$_6$Sn$_6$, where the concept of Weyl points is not easily applicable.

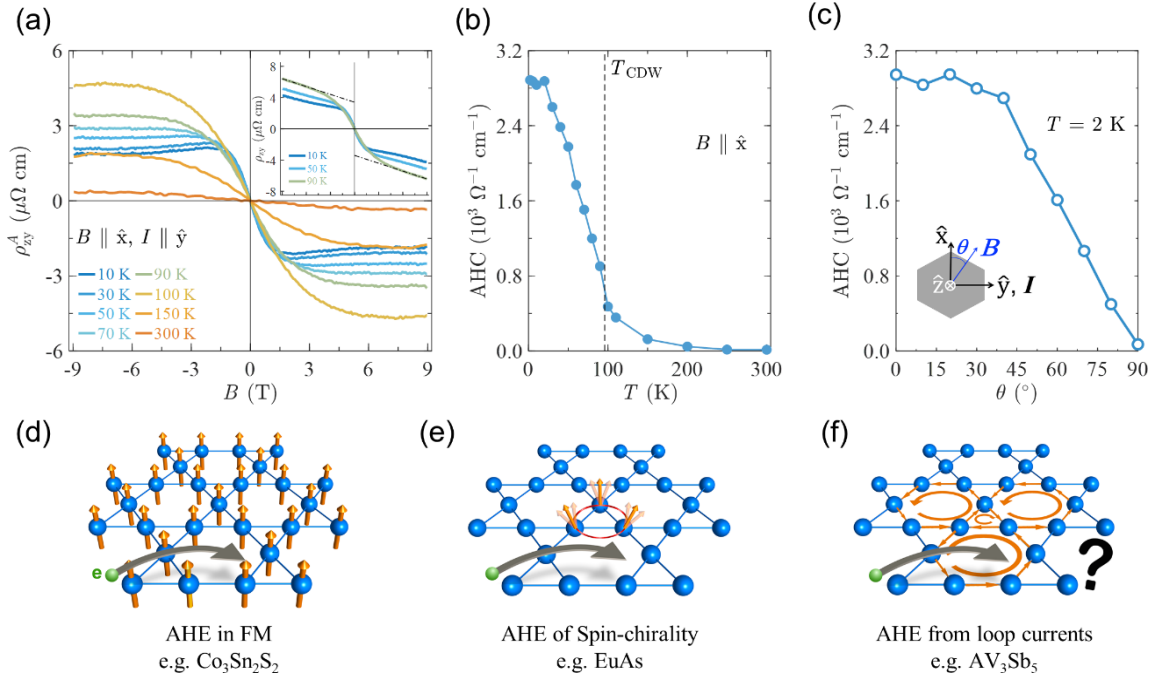

FIG. 5. Anomalous Hall-like behavior in ScV$_6$Sn$_6$. (a) Field-dependent measured anomalous behavior of Hall resistivity, $\rho_{zy}$ at various temperatures (inset) and extracted Hall resistivity $\rho_{zy}^A$. (b) Temperature and (c) angular dependent corresponding anomalous Hall like conductivity. Electrons picking up anomalous velocity from the different sources and their typical example. (d) Non-zero Berry phase from ferromagnetic spins. (e) Topological orbital moment due to spins chirality. (f) Formation of local loop currents.

In conclusion, with the help of field-dependent electrical transports and theoretical calculations, we have revealed the fermiology of ScV$_6$Sn$_6$. Along with a large FS, the BZ is comprised of the several other small electron FSs, except for the one tiny hole FS. The small FSs exhibit SdH oscillations, resulting in the small effective mass, and the symmetric and asymmetric nature of the FSs for electron and hole, respectively. The electron Dirac band located at the $K$ point gives rise to the non-zero Berry phase, proving that it is a non-trivial topological FS. Furthermore, the non-linear Hall, which resembles the anomalous Hall, does not fit well with the two-carrier model. Finding the microscopic origins for the anomalous Hall-like signal remains an important issue that requires further theoretical and experimental investigations.

Note added. During the preparation of our work, a similar work including anomalous Hall-like behavior in ScV6Sn6 was found [59].

**ACKNOWLEDGMENTS**

This work was financially supported by the Deutsche Forschungsgemeinschaft (DFG) under SFB1143 (project no. 247310070), the Würzburg-Dresden Cluster of Excellence on Complexity and Topology in Quantum Matter—ct.qmat (EXC 2147, project no. 390858490) and the QUAST-FOR5249-449872909. S.R. thanks the Alexander von Humboldt Foundation for a fellowship.

**APPENDIX A: MAGNETIC AND ELECTRICAL MEASUREMENTS**

Temperature-dependent susceptibility (2–300 K) under various magnetic fields was measured in zero field cooled and field cooled configuration in a Magnetic Properties Measurement System (MPMS, Quantum Design Inc.) equipped with a Superconducting Quantum Interference Device - Vibrating Sample Magnetometer (SQUID-VSM) option. Temperature-dependent four-probe transverse and Hall resistivities were measured on a standard rotating sample holder in Physical Property Measurement System (PPMS, Quantum Design Inc.). The alternative current transport (ACT) option was used for measurement with an excitation current of 1.5 mA at a fixed frequency of 93 Hz. Data were collected in the temperature range of 2–300 K, and the magnetic field range of -9 to 9 T in sweep mode.

**APPENDIX B: ELECTRONIC BAND STRUCTURE CALCULATIONS**

Ab initio calculations were performed using density functional theory (DFT) implemented in the Vienna Ab initio Simulation Package (VASP) [60,61]. The projector augmented wave method [62] and the generalized gradient approximation (GGA) with Perdew-Burke-Ernzerhof (PBE) exchange-correlation functional [63] were used for the calculation. Here, we used refined lattice parameters for the band structure calculations, including the pristine structure and the CDW structure. A 15 × 15 × 7 *k*-mesh was used to sample the Brillouin zone of the pristine structure, with an energy cut-off of 500 eV. For the CDW structure, an 8 × 8 × 8 *k*-mesh was used, with an energy cut-off of 450 eV. The CDW band structures were unfolded using the VASPKIT toolkit [64]. To obtain the quantum oscillation frequencies, we constructed a Wannier tight-binding Hamiltonian using the Wannier90 code [65], including Sc-3*d*, V-3*d*, and Sn-5*p* orbitals.

**References:**

[1] L.-K. Lim, J.-N. Fuchs, F. Piéchon, and G. Montambaux, Dirac points emerging from flat bands in Lieb-kagome lattices, Phys. Rev. B 101, 045131 (2020).

[2] M. Li, Q. Wang, G. Wang, Z. Yuan, W. Song, R. Lou, Z. Liu, Y. Huang, Z. Liu, H. Lei et al., Dirac cone, flat band and saddle point in kagome magnet $YMn_6Sn_6$, Nat. Commun. 12, 3129 (2021).

[3] W.-S. Wang, Z.-Z. Li, Y.-Y. Xiang, and Q.-H. Wang, Competing electronic orders on kagome lattices at van Hove filling, Phys. Rev. B 87, 115135 (2013).

[4] Y. Hu, X. Wu, Y. Yang, S. Gao, N. C. Plumb, A. P. Schnyder, W. Xie, J. Ma, and M. Shi, Tunable topological Dirac surface states and van Hove singularities in kagome metal $GdV_6Sn_6$, Sci. Adv. 8, eadd2024 (2022).

[5] B. R. Ortiz, S. M. L. Teicher, Y. Hu, J. L. Zuo, P. M. Sarte, E. C. Schueller, A. M. M. Abeykoon, M. J. Krogstad, S. Rosenkranz, R. Osborn et al., $CsV_3Sb_5$: A $Z_2$ topological kagome metal with a superconducting ground state, Phys. Rev. Lett. 125, 247002 (2020).

[6] H.-M. Guo and M. Franz, Topological insulator on the kagome lattice, Phys. Rev. B 80, 113102 (2009).

[7] B. R. Ortiz, L. C. Gomes, J. R. Morey, M. Winiarski, M. Bordelon, J. S. Mangum, I. W. H. Oswald, J. A. Rodriguez-Rivera, J. R. Neilson, S. D. Wilson et al., New kagome prototype materials: Discovery of $KV_3Sb_5$, $RbV_3Sb_5$, and $CsV_3Sb_5$, Phys. Rev. Mater. 3, 094407 (2019).

[8] J.-X. Yin, W. Ma, T. A. Cochran, X. Xu, S. S. Zhang, H.-J. Tien, N. Shumiya, G. Cheng, K. Jiang, B. Lian et al., Quantum-limit Chern topological magnetism in $TbMn_6Sn_6$, Nature (London) 583, 533 (2020).

[9] L. Zheng, Z. Wu, Y. Yang, L. Nie, M. Shan, K. Sun, D. Song, F. Yu, J. Li, D. Zhao et al., Emergent charge order in pressurized kagome superconductor $CsV_3Sb_5$, Nature (London) 611, 682 (2022).

[10] X. Wu, T. Schwemmer, T. Müller, A. Consiglio, G. Sangiovanni, D. Di Sante, Y. Iqbal, W. Hanke, A. P. Schnyder, M. M. Denner et al., Nature of unconventional pairing in the kagome superconductors $AV_3Sb_5$ (A = K, Rb, Cs), Phys. Rev. Lett. 127, 177001 (2021).

[11] X. Teng, L. Chen, F. Ye, E. Rosenberg, Z. Liu, J.-X. Yin, Y.-X. Jiang, J. S. Oh, M. Z. Hasan, K. J. Neubauer et al., Discovery of charge density wave in a kagome lattice antiferromagnet, Nature (London) 609, 490 (2022).

[12] H. Tan, Y. Liu, Z. Wang, and B. Yan, Charge density waves and electronic properties of superconducting kagome metals, Phys. Rev. Lett. 127, 046401 (2021).

[13] X. Feng, K. Jiang, Z. Wang, and J. Hu, Chiral flux phase in the kagome superconductor $AV_3Sb_5$, Sci. Bull. 66, 1384 (2021).

[14] F. H. Yu, T. Wu, Z. Y. Wang, B. Lei, W. Z. Zhuo, J. J. Ying, and X. H. Chen, Concurrence of anomalous Hall effect and charge density wave in a superconducting topological kagome metal, Phys. Rev. B 104, L041103 (2021).

[15] S.-Y. Yang, Y. Wang, B. R. Ortiz, D. Liu, J. Gayles, E. Derunova, R. Gonzalez-Hernandez, L. Šmejkal, Y. Chen, S. S. P. Parkin et al., Giant, unconventional anomalous Hall effect in the metallic frustrated magnet candidate, $KV_3Sb_5$, Sci. Adv. 6, eabb6003 (2020).

[16] C. Mielke, D. Das, J.-X. Yin, H. Liu, R. Gupta, Y.-X. Jiang, M. Medarde, X. Wu, H. C. Lei, J. Chang et al., Time-reversal symmetry-breaking charge order in a kagome superconductor, Nature (London) 602, 245 (2022).

[17] H. Li, H. Zhao, B. R. Ortiz, T. Park, M. Ye, L. Balents, Z. Wang, S. D. Wilson, and I. Zeljkovic, Rotation symmetry breaking in the normal state of a kagome superconductor $KV_3Sb_5$, Nat. Phys. 18, 265 (2022).

[18] C. Guo, C. Putzke, S. Konyzheva, X. Huang, M. Gutierrez-Amigo, I. Errea, D. Chen, M. G. Vergniory, C. Felser, M. H. Fischer et al., Switchable chiral transport in charge-ordered kagome metal $CsV_3Sb_5$, Nature (London) 611, 461 (2022).

[19] X. Teng, J. S. Oh, H. Tan, L. Chen, J. Huang, B. Gao, J.-X. Yin, J.-H. Chu, M. Hashimoto, D. Lu et al., Magnetism and charge density wave order in kagome FeGe, Nat. Phys. 19, 814 (2023).

[20] H. W. S. Arachchige, W. R. Meier, M. Marshall, T. Matsuoka, R. Xue, M. A. McGuire, R. P. Hermann, H. Cao, and D. Mandrus, Charge density wave in kagome lattice intermetallic $ScV_6Sn_6$, Phys. Rev. Lett. 129, 216402 (2022).

[21] X. Zhang, J. Hou, W. Xia, Z. Xu, P. Yang, A. Wang, Z. Liu, J. Shen, H. Zhang, X. Dong et al., Destabilization of the charge density wave and the absence of superconductivity in ScV6Sn6 under high pressures up to 11 GPa, Materials 15, 7372 (2022).

[22] Y. Hu, J. Ma, Y. Li, D. J. Gawryluk, T. Hu, J. Teyssier, V. Multian, Z. Yin, Y. Jiang, S. Xu et al., Phonon promoted charge density wave in topological kagome metal $ScV_6Sn_6$, arXiv:2304.06431.

[23] T. Hu, H. Pi, S. Xu, L. Yue, Q. Wu, Q. Liu, S. Zhang, R. Li, X. Zhou, J. Yuan et al., Optical spectroscopy and band structure calculations of the structural phase transition in the vanadium-based kagome metal ScV6Sn6, Phys. Rev. B 107, 165119 (2023).

[24] A. Korshunov, H. Hu, D. Subires, Y. Jiang, D. Cǎlugǎru, X. Feng, A. Rajapitamahuni, C. Yi, S. Roychowdhury, M. G. Vergniory et al., Softening of a flat phonon mode in the kagome ScV6Sn6, Nat. Commun. 14, 6646 (2023).

[25] Y. Gu, E. T. Ritz, W. R. Meier, A. Blockmon, K. Smith, R. P. Madhogaria, S. Mozaffari, D. Mandrus, T. Birol, and J. L. Musfeldt, Phonon mixing in the charge density wave state of ScV6Sn6, npj Quantum Mater. 8, 58 (2023).

[26] D. Di Sante, C. Bigi, P. Eck, S. Enzner, A. Consiglio, G. Pokharel, P. Carrara, P. Orgiani, V. Polewczyk, J. Fujii et al., Flat band separation and robust spin Berry curvature in bilayer kagome metals, Nat. Phys. 19, 1135 (2023).

[27] S. Cheng, Z. Ren, H. Li, J. Oh, H. Tan, G. Pokharel, J. M. DeStefano, E. Rosenberg, Y. Guo, Y. Zhang et al., Nanoscale visualization and spectral fingerprints of the charge order in ScV6Sn6 distinct from other kagome metals, arXiv:2302.12227.

[28] S. Lee, C. Won, J. Kim, J. Yoo, S. Park, J. Denlinger, C. Jozwiak, A. Bostwick, E. Rotenberg, R. Comin et al., Nature of charge density wave in kagome metal ScV6Sn6, arXiv:2304.11820.

[29] Z. Guguchia, D. J. Gawryluk, S. Shin, Z. Hao, C. Mielke, III, D. Das, I. Plokhikh, L. Liborio, K. Shenton, Y. Hu et al., Hidden magnetism uncovered in charge ordered bilayer kagome material ScV6Sn6, Nat. Commun. 14, 7796 (2023).

[30] See Supplemental Material at http://link.aps.org/supplemental/10.1103/PhysRevB.xx.xxxxxx for growth method, structural and chemical component characterization of single- crystal ScV6Sn6; magnetic properties; additional electromag- netic transport properties; and calculated electronic band structures.

[31] J. Rodríguez-Carvajal, Recent advances in magnetic structure determination by neutron powder diffraction, Physica B 192, 55 (1993).

[32] G. Pokharel, S. M. L. Teicher, B. R. Ortiz, P. M. Sarte, G. Wu, S. Peng, J. He, R. Seshadri, and S. D. Wilson, Electronic properties of the topological kagome metals YV6Sn6 and GdV6Sn6, Phys. Rev. B 104, 235139 (2021).

[33] C. Shekhar, A. K. Nayak, Y. Sun, M. Schmidt, M. Nicklas, I. Leermakers, U. Zeitler, Y. Skourski, J. Wosnitza, Z. Liu et al., Extremely large magnetoresistance and ultrahigh mobility in the topological weyl semimetal candidate NbP, Nat. Phys. 11, 645 (2015).

[34] D. Shoenberg, Magnetic Oscillations in Metals (Cambridge University Press, Cambridge, 1984).

[35] B. R. Ortiz, S. M. L. Teicher, L. Kautzsch, P. M. Sarte, N. Ratcliff, J. Harter, J. P. C. Ruff, R. Seshadri, and S. D. Wilson, Fermi surface mapping and the nature of charge-density-wave order in the kagome superconductor CsV3Sb5, Phys. Rev. X 11, 041030 (2021).

[36] Y. Fu, N. Zhao, Z. Chen, Q. Yin, Z. Tu, C. Gong, C. Xi, X. Zhu, Y. Sun, K. Liu et al., Quantum transport evidence of topological band structures of kagome superconductor CsV3Sb5, Phys. Rev. Lett. 127, 207002 (2021).

[37] K. Shrestha, M. Shi, B. Regmi, T. Nguyen, D. Miertschin, K. Fan, L. Z. Deng, N. Aryal, S.-G. Kim, D. E. Graf et al., High quantum oscillation frequencies and nontriv- ial topology in kagome superconductor KV3Sb5 probed by torque magnetometry up to 45 T, Phys. Rev. B 107, 155128 (2023).

[38] Y. Wang, Z. Chen, Y. Nie, Y. Zhang, Q. Niu, G. Zheng, X. Zhu, W. Ning, and M. Tian, Nontrivial fermi surface topology and large anomalous Hall effect in the kagome superconductor RbV3Sb5 Phys. Rev. B 108, 035117 (2023).

[39] Y. Ando, Topological insulator materials, J. Phys. Soc. Jpn. 82, 102001 (2013).

[40] A. A. Taskin, S. Sasaki, K. Segawa, and Y. Ando, Manifestation of topological protection in transport properties of epitaxial Bi2Se3 thin films, Phys. Rev. Lett. 109, 066803 (2012).

[41] F.-X. Xiang, X.-L. Wang, M. Veldhorst, S.-X. Dou, and M. S. Fuhrer, Observation of topological transition of Fermi surface from a spindle torus to a torus in bulk Rashba spin-split BiTeCl, Phys. Rev. B 92, 035123 (2015).

[42] H. Murakawa, M. S. Bahramy, M. Tokunaga, Y. Kohama, C. Bell, Y. Kaneko, N. Nagaosa, H. Y. Hwang, and Y. Tokura, Detection of Berry's phase in a bulk Rashba semiconductor, Science 342, 1490 (2013).

[43] D.-X. Qu, Y. S. Hor, J. Xiong, R. J. Cava, and N. P. Ong, Quantum oscillations and Hall anomaly of surface states in the topological insulator Bi2Te3, Science 329, 821 (2010).

[44] X. Xu, X. Wang, T. A. Cochran, D. S. Sanchez, G. Chang, I. Belopolski, G. Wang, Y. Liu, H.-J. Tien, X. Gui et al., Crys- tal growth and quantum oscillations in the topological chiral semimetal CoSi, Phys. Rev. B 100, 045104 (2019).

[45] N. Huber, K. Alpin, G. L. Causer, L. Worch, A. Bauer, G. Benka, M. M. Hirschmann, A. P. Schnyder, C. Pfleiderer, and M. A. Wilde, Network of topological nodal planes, multifold degeneracies, and weyl points in CoSi, Phys. Rev. Lett. 129, 026401 (2022).

[46] M. N. Ali, L. M. Schoop, C. Garg, J. M. Lippmann, E. Lara, B. Lotsch, and S. S. P. Parkin, Butterfly magnetoresistance, quasi-2D Dirac Fermi surface and topological phase transition in ZrSiS, Sci. Adv. 2, e1601742 (2016).

[47] N. Bansal, Y. S. Kim, M. Brahlek, E. Edrey, and S. Oh, Thickness-independent transport channels in topological insu- lator Bi2Se3 thin films, Phys. Rev. Lett. 109, 116804 (2012).

[48] F.C.Chen,Y.Fei,S.J.Li,Q.Wang,X.Luo,J.Yan,W.J. Lu, P. Tong, W. H. Song, X. B. Zhu et al., Temperature-induced lifshitz transition and possible excitonic instability in ZrSiSe, Phys. Rev. Lett. 124, 236601 (2020).

[49] J. Xu, Y. Wang, S. E. Pate, Y. Zhu, Z. Mao, X. Zhang, X. Zhou, U. Welp, W.-K. Kwok, D. Y. Chung et al., Unreliability of two-band model analysis of magnetoresistivities in unveil- ing temperature-driven Lifshitz transition, Phys. Rev. B 107, 035104 (2023).

[50] Y. L. Wang, L. R. Thoutam, Z. L. Xiao, J. Hu, S. Das, Z. Q. Mao, J. Wei, R. Divan, A. Luican-Mayer, G. W. Crabtree et al., Origin of the turn-on temperature behavior in WTe2, Phys. Rev. B 92, 180402 (2015).

[51] X. Huang, L. Zhao, Y. Long, P. Wang, D. Chen, Z. Yang, H. Liang, M. Xue, H. Weng, Z. Fang et al., Observation of the chiral-anomaly-induced negative magnetoresistance in 3D weyl semimetal TaAs, Phys. Rev. X 5, 031023 (2015).

[52] N. Kumar, Y. Sun, N. Xu, K. Manna, M. Yao, V. Süss, I. Leermakers, O. Young, T. Förster, M. Schmidt et al., Extremely high magnetoresistance and conductivity in the type- II weyl semimetals WP2 and MoP2, Nat. Commun. 8, 1642 (2017).

[53] J. A. Voerman, L. Mulder, J. C. De Boer, Y. Huang, L. M. Schoop, C. Li, and A. Brinkman, Origin of the butterfly magne- toresistance in ZrSiS, Phys. Rev. Mater. 3, 084203 (2019).

[54] K. Manna, L. Muechler, T.-H. Kao, R. Stinshoff, Y. Zhang, J. Gooth, N. Kumar, G. Kreiner, K. Koepernik, R. Car et al., From colossal to zero: Controlling the anomalous Hall effect in magnetic Heusler compounds via Berry curvature design, Phys. Rev. X 8, 041045 (2018).

[55] D. S. Bouma, Z. Chen, B. Zhang, F. Bruni, M. E. Flatté, A. Ceballos, R. Streubel, L.-W. Wang, R. Q. Wu, and F. Hellman, Itinerant ferromagnetism and intrinsic anomalous Hall effect in amorphous iron-germanium, Phys. Rev. B 101, 014402 (2020).

[56] R. Khasanov, D. Das, R. Gupta, C. Mielke, M. Elender, Q. Yin, Z. Tu, C. Gong, H. Lei, E. T. Ritz et al., Time-reversal symmetry broken by charge order in CsV3Sb5, Phys. Rev. Res. 4, 023244 (2022).

[57] M. H. Christensen, T. Birol, B. M. Andersen, and R. M. Fernandes, Loop currents in CsV3Sb5 kagome metals: Multi- polar and toroidal magnetic orders, Phys. Rev. B 106, 144504 (2022).

[58] T. Liang, J. Lin, Q. Gibson, S. Kushwaha, M. Liu, W. Wang, H. Xiong, J. A. Sobota, M. Hashimoto, P. S. Kirchmann et al., Anomalous Hall effect in ZrTe5, Nat. Phys. 14, 451 (2018).

[59] S. Mozaffari, W. R. Meier, R. P. Madhogaria, N. Peshcherenko, S.-H. Kang, J. W. Villanova, H. W. S. Arachchige, G. Zheng, Y. Zhu, K.-W. Chen et al., Universal sublinear resistivity in vanadium kagome materials hosting charge density waves, arXiv:2305.02393.

[60] G. Kresse and J. Furthmüller, Efficiency of ab-initio total energy calculations for metals and semiconductors using a plane-wave basis set, Comput. Mater. Sci. 6, 15 (1996).

[61] G. Kresse and J. Furthmüller, Efficient iterative schemes for ab initio total-energy calculations using a plane-wave basis set, Phys. Rev. B 54, 11169 (1996).

[62] P. E. Blöchl, Projector augmented-wave method, Phys. Rev. B 50, 17953 (1994).

[63] J. P. Perdew, K. Burke, and M. Ernzerhof, Generalized gradient approximation made simple, Phys. Rev. Lett. 77, 3865 (1996).

[64] V. Wang, N. Xu, J.-C. Liu, G. Tang, and W.-T. Geng, VASPKIT: A user-friendly interface facilitating high- throughput computing and analysis using VASP code, Comput. Phys. Commun. 267, 108033 (2021).

[65] G. Pizzi, V. Vitale, R. Arita, S. Blügel, F. Freimuth, G. Géranton, M. Gibertini, D. Gresch, C. Johnson, T. Koretsune et al., Wannier90 as a community code: New features and applications, J. Phys.: Condens. Matter 32, 165902 (2020).

## Supplementary information

**Single crystal growth**

High-quality single crystals of ScV$_6$Sn$_6$ were grown by the flux method [1]. High-purity starting elements Sc (99.99 %, Alfa Aesar), V (99.98 %, Alfa Aesar), and Sn (99.999 %, Alfa Aesar) were placed in an alumina crucible in a molar ratio of 1 : 10 : 60 and then sealed in an evacuated quartz tube. The tube was then slowly heated to 1373 K, maintained for 10 h, and cooled down to 973 K over 400 h. Hexagonal shape crystals with silvery surfaces and a typical size of 2 × 2 × 1 mm$^3$ were obtained after centrifugation. Hall bars were cut from the crystals and polished before measurements were taken. The crystal orietations were determined by using the Laue diffractometer equipped with the Mo- X-ray source.

**Crystal structure characterization**

As grown crystals of ScV$_6$Sn$_6$ were grinded into the fine powders to perform the powder X-ray diffraction (PXRD) experiment. As shown in Fig. S1, the observed diffraction pattern at room temperature can be well refined using the *P*6/*mmm* space group by the FullProf program [2]. After the refinement, the lattice parameters are $a = b = 5.47332(5)$ Å and $c = 9.17248(8)$ Å, which are very close to the lattice parameters of ScV$_6$Sn$_6$ reported in the previous report [1]. Only a tiny impurity peak is observed due to the Sn flux as many pieces were grind together for PXRD.

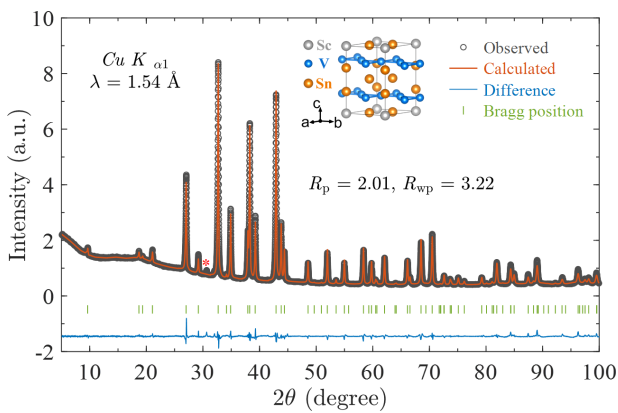

FIG. S1. Crystal structure, powder x-ray diffraction pattern. The powder x-ray diffraction pattern of grinded ScV$_6$Sn$_6$ single crystals. The inset is a schematic picture of the crystal structure in 3D view. The red asterisk is marked for impurity, possibly Sn.

**Electromagnetic measurements**

Temperature-dependent susceptibility (2–300 K) under various magnetic fields was measured in zero field cooled (ZFC) and field cooled (FC) configuration in an MPMS SQUID-VSM (Quantum Design). Temperature-dependent four-probe transverse and Hall resistivities were measured on a standard rotating sample holder in PPMS (Quantum Design). The alternative current transport (ACT) option was used for measurement with an excitation current of 1.5 mA at a fixed frequency of 93 Hz. Data were collected in the temperature range of 2–300 K, and the magnetic field range of -9 to 9 T in sweep mode.

For the magnetic property measurements of $ScV_6Sn_6$, we selected flux-free crystals and a typical optical image as shown in the inset of Fig. S2 (a). For ease of further reference, we marked the crystal axes in the Cartesian coordinates $\hat{x}, \hat{y}, \hat{z}$, which are correspond to ($2\bar{1}\bar{1}0$), ($01\bar{1}0$), (0001) crystallographic directions in the hexagonal crystal structure. The isomagnetic and isothermal magnetic properties were measured by applying a field along $\hat{x}$ and $\hat{z}$, the measured data are shown in Figs. S2 (a) (b) and S2 (c) (d), respectively. The magnetization decreases linearly with temperature and then suddenly drops at 92 K. The sudden decrease marks the charge density wave (CDW) transition temperature, $T_{CDW}$. Noticeably, this magnetic above $T_{CDW}$ does not follow the Curie-Weiss paramagnetic characteristic as its sister compound $YV_6Sn_6$ [3] follows. After 50 K which is far below the $T_{CDW}$, the magnetization starts to increase with decreasing temperature and behaves like a week Pauli paramagnetic behavior [4]. This is also a distinct behavior from the well-known $AV_3Sb_5$ (A= K, Rb, Cs) family, which follows the Curie-Weiss law with a small effective moment (0.22 $\mu_B$ / V atom for $KV_3Sb_5$) [5]. From Fig. S2 (b), the field dependent magnetization varies linearly at all measured temperatures and does not show any nonlinear behavior. When magnetic field is applied along $\hat{z}$, as shown in Fig. S1 (c) and (d), the magnetization is slightly larger than along $\hat{x}$ but follows almost the same behavior. Remarkably, the dip induced by the magnetic field is observed at about 250 K when the magnetic field is aligned along $\hat{z}$. However, the reason for the dip is still unclear and may need further attention.

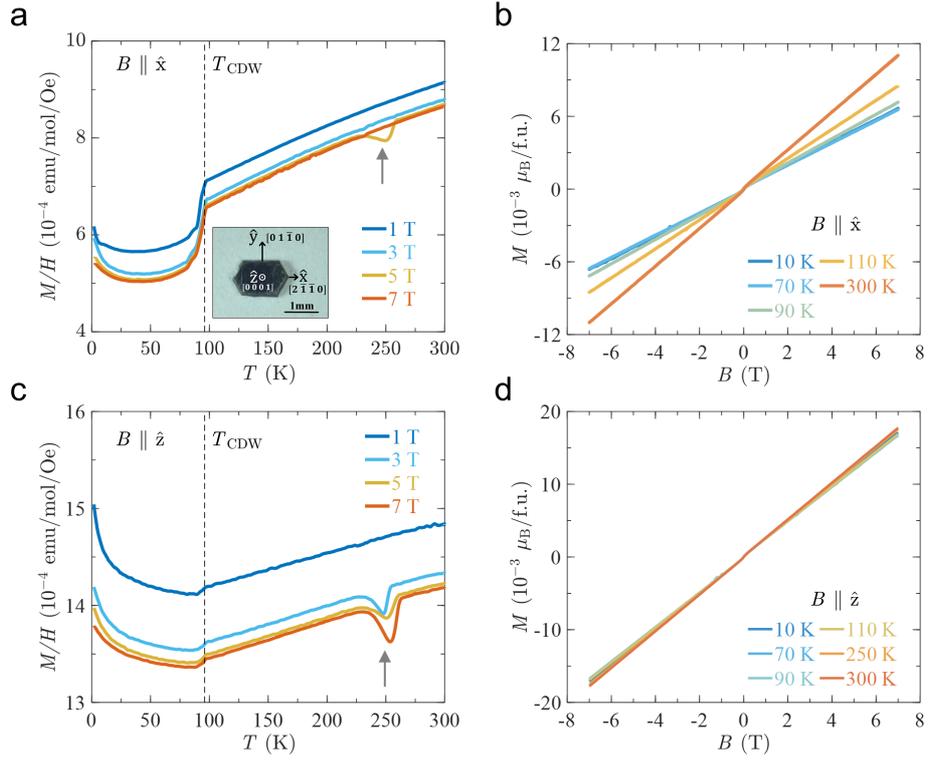

FIG. S2. Magnetic measurements. (a), and (c), Temperature-dependent magnetization at several magnetic fields $B \parallel \hat{x}$ and $B \parallel \hat{z}$, respectively. (b), and (d), Isothermal magnetization as a function of magnetic fields $B \parallel \hat{x}$ and $B \parallel \hat{z}$, respectively for different temperatures.

**Additional electronic transport measurements and analyses**

Figure S3 shows the additional measurement of electronic transport properties when magnetic field is applied along $\hat{x}$ and electric current is injected along $\hat{y}$. The temperature dependent resistivity shows a sharp drop at around $T_{CDW}$ and follows almost the same behavior with different fields as shown in Fig. S3 (a). The magnetoresistance (MR = $(\rho(B)/\rho(0) – 1) \times 100\%$), shown in Fig. S3 (b), follows a parabolic dependence in the low magnetic field range, but it becomes linear with further field increase. The concentration and mobility of the electron-type carrier extracted from the Hall resistivity in the high field range show an obvious change around $T_{CDW}$, which may be due to the phonon softening. A dramatic sign change of the carrier concentration, which appears only at the CDW phase transition, implies a strong interaction between charge carrier and lattice dynamics. The values of concentration and mobility at 2K extracted from the high field slope of the Hall resistivity are $2.36 \times 10^{21}$ cm$^{-3}$ and 265 cm$^2$ V$^{-1}$ s$^{-1}$, respectively, as shown in Fig. S3(c). The charge carriers decrease slightly with increasing temperature, while the mobility decreases more rapidly with increasing temperature up to the CDW transition. Fig. S3(d) shows the Hall resistivity as function of the rotating angle from $B \parallel \hat{x}$ to $B \parallel \hat{y}$, i.e., from being perpendicular to the current to parallel to the current, which also shows the anomalous Hall effect (AHE) like non-linear behavior.

Then we changed the current direction and now the current is along $\hat{z}$ but the field is the same as along $\hat{x}$. The measured data are displayed in Fig. S4 and show similar behavior to the $I \parallel \hat{y}$ and $B \parallel \hat{x}$ measurement geometry, differing only in magnitude. For example, the $\rho_{zz}$ shows a stronger dependence on the magnetic field, electron type charge carrier as expected from the electronic band structure. Additionally, the $\rho_{zz}$ (Fig. S4d) as a function of angle dependent at various magnetic fields have an obvious twofold symmetry when the field is rotated in a full circle, following a conventional trigonometric dependence. From our results, the magnetization is almost negligible in ScV$_6$Sn$_6$ as shown in Fig. S2. However, the Hall measurements exhibit a similar feature of AHE, which is seen up to the CDW phase transition temperature.

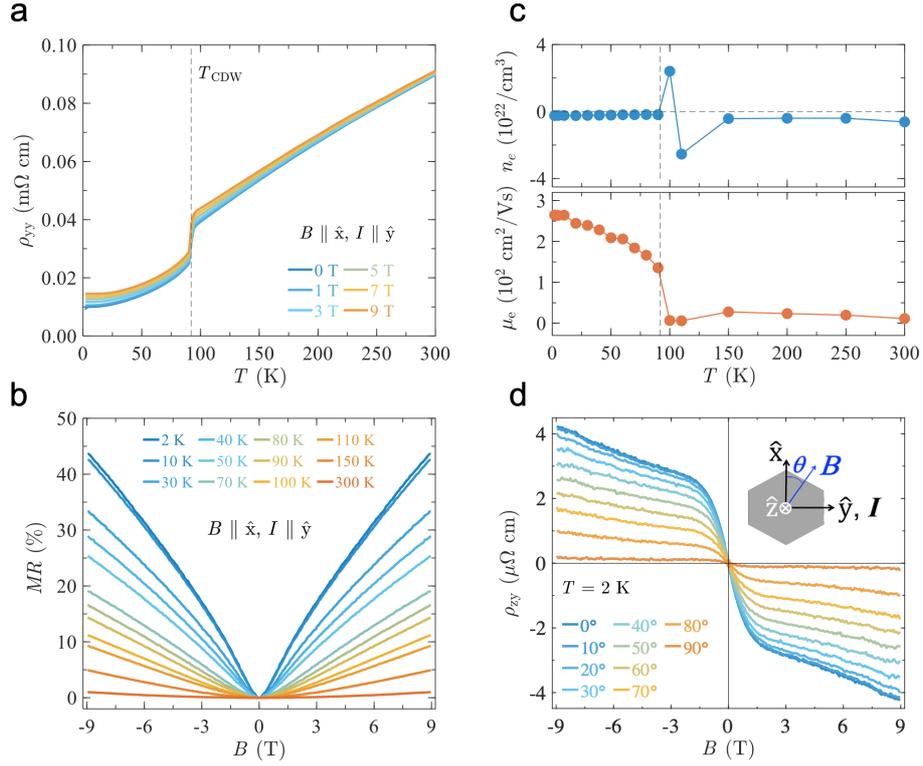

FIG. S3. Resistivity, MR, carrier density and mobility, Hall resistivity. (a) Temperature-dependent resistivity $\rho_{yy}$ at various fields when $I \parallel \hat{y}$, $B \parallel \hat{x}$. (b) Magnetoresistance as functions of magnetic field at various temperatures. (c) Estimated carrier concentration and mobility as functions of temperature. The sudden change is related to CDW phase transition, which is also highlighted by the vertical black dashed line. (d) Hall resistivity as functions of magnetic field at different rotating angles. The field is rotated from the $\hat{x}$ to the $\hat{y}$ directions as shown in the inset.

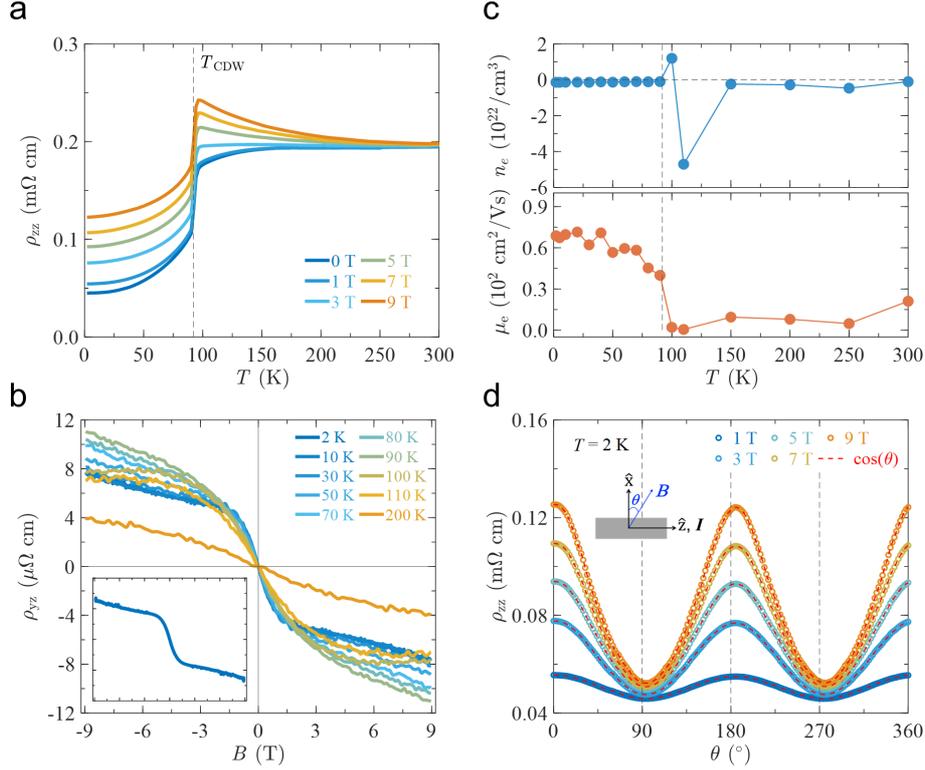

FIG. S4. Resistivity, anomalous Hall, carrier density and mobility, angular resistivity. (a) Temperature-dependent resistivity $\rho_{zz}$ when $I \parallel \hat{z}$, $B \parallel \hat{x}$ at different magnetic fields. (b) Hall resistivity $\rho_{yz}$ as functions of magnetic field at different temperatures. (c) Estimated carrier concentration and mobility as functions of temperature. (d) Angle-dependent longitudinal resistivity $\rho_{zz}$ at 2 K for several magnetic fields. The red dashed lines are cosine function fit. The inset shows that the field is rotated from $\hat{x}$ to $\hat{z}$.

A nonlinear Hall signal in nonmagnetic systems is primarily attributed to the presence of multiple bands. Such Hall is described by the standard two-band model, as

$$\rho_{zy}(B) = \frac{1}{e} \frac{\mu_1^2 \mu_2^2 (n_1+n_2) B^3 + (\mu_1^2 n_1 + \mu_2^2 n_2) B}{\mu_1^2 \mu_2^2 (n_1+n_2)^2 B^2 + (\mu_1 n_1 + \mu_2 n_2)^2} \tag{S1}$$

where $\mu_1, \mu_2$ are the mobilities and $n_1, n_2$ are the density of the carriers, respectively. For a more reliable analysis, the Hall conductivity, $\sigma_{yz} = \frac{\rho_{zy}}{\rho_{zz}\rho_{yy} + \rho_{zy}\rho_{yz}} \approx \frac{\rho_{zy}}{\rho_{zz}\rho_{yy}}$ and the longitudinal conductivity, $\sigma_{yy} = \frac{\rho_{zz}}{\rho_{zz}\rho_{yy} + \rho_{zy}\rho_{yz}} \approx \frac{1}{\rho_{yy}}$, when $\rho_{zz}\rho_{yy} \gg \rho_{zy}\rho_{yz}$, are estimated from the measured Hall and longitudinal resistivities. Then, the following conventional relations of the two-band model for Hall and longitudinal conductivities are fitted [6–8]:

$$\sigma_{yz}(B) = eB\left(\frac{n_1 \mu_1^2}{1+\mu_1^2 B^2} + \frac{n_2 \mu_2^2}{1+\mu_2^2 B^2}\right), \tag{S2}$$

$$\sigma_{yy}(B) = e\left(\frac{n_1 \mu_1}{1+\mu_1^2 B^2} + \frac{n_2 \mu_2}{1+\mu_2^2 B^2}\right). \tag{S3}$$

The fitted results of $\sigma_{yz}$ and $\sigma_{yy}$ are summarized in Fig. S5. From Figs. S5 (a) and (b), the $\sigma_{yz}$ and $\sigma_{yy}$ are well fitted using Eq. S2 and Eq. S3, respectively. However, the fitted parameters are completely different from each other: the fit of $\sigma_{yz}$ indicates electron carriers only, while the fit of $\sigma_{yy}$ gives both electron and hole carriers, indicating a clear disparity between these two fits. Usually, both fits reflect the same parameters, conforming the applicability of the two-band model, as reported in several typical multiband systems [8,9]. We also examined the $\sigma_{yz}$ and $\sigma_{yy}$ fits by considering the same initial parameters from Figs. S5 (a) and (b) in the two-band model. In this way, the fits are the worst and do not converge as shown in Figs. 5 (c) and (d). Therefore, the two-band model may not be the correct mechanism to describe the nonlinear Hall behavior in ScV$_6$Sn$_6$.

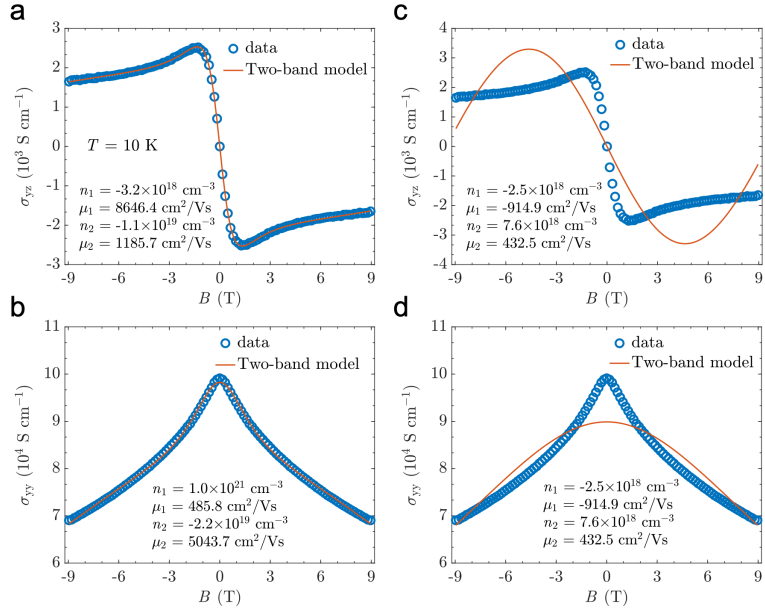

FIG. S5. Two-band model fitting. Individual fitting of (a) Hall conductivity by Eq. S2 and (b) longitudinal conductivity by Eq. S3 at $T$ = 10 K, respectively. (c-d) Simultaneously fitting of Hall and longitudinal conductivities.

**Additional Shubnikov de-Haas oscillations analyses**

The other important physical property, the Shubnikov–de Haas (SdH) quantum oscillation, is also observed in the high-quality $ScV_6Sn_6$ crystals, as shown in Fig. S6. We present the first derivative of $\rho_{zz}$ in Fig. S6b to show the SdH oscillation more clearly. The large magnetoresistance (170% at 2 K and 9 T) and the SdH oscillation indicate the high quality of our crystals. Fig. S6c shows the frequency dependence calculated by Fast Fourier Transformation (FFT), which gives three frequencies at 10, 28, and 50 T labeled as $\alpha$, $\beta$, and $\gamma$ with respect to three Fermi pockets as discussed in the main text. Due to the multiple frequency as we mentioned in the main text, it is difficult to estimate the parameters related to the Fermi surfaces (FSs) from the whole FFT spectrum. Thus, we extracted the frequency components for further analysis by using bandwidth pass as shown in Fig. S6d [10]. However, by using FFT filtering process the oscillation decay as increasing magnetic field (i.e. decreasing $1/B$), as shown in Fig. S6e and Fig. S7a for reconstructed $\beta$ - and $\gamma$-pocket, respectively, suggesting the presence of beating patterns which influence the Berry's phase analysis [11,12]. Thus, the extracted result is relatively unreliable. We also directly analyzed the angle dependent SdH curves for some curves that have only one frequency, as shown in Fig. S7b – f. the estimated Berry's phases are listed and discussed in Fig. 2g in the main text. The so-called Dingle plot is also presented in Fig. S8 for $\gamma$ pocket when magnetic field is parallel to $\hat{z}$-axis, giving $T_D$ = 2.43 K and $\tau_q = 5 \times 10^{13}$ s for $F_\gamma$. The other estimated parameters of both $\alpha$, $\beta$, and $\gamma$ FS pockets are listed in Table S1.

The magnetoresistance (MR) for the rotating magnetic field from $\hat{x}$ to $\hat{z}$ and $\hat{x}$ to $\hat{y}$ are shown in Fig. S9. At $T$ = 2 K, the oscillation of $\Delta\rho_{zz}$ shows an obvious change when the field is rotated in the $\hat{x}$ - $\hat{z}$ plane, illustrating an anisotropic FS. Contrastingly, when magnetic field is rotating in $\hat{x}$ - $\hat{y}$ plane, the oscillation spectrum is almost the same, indicating that the FS is almost isotropic in this plane.

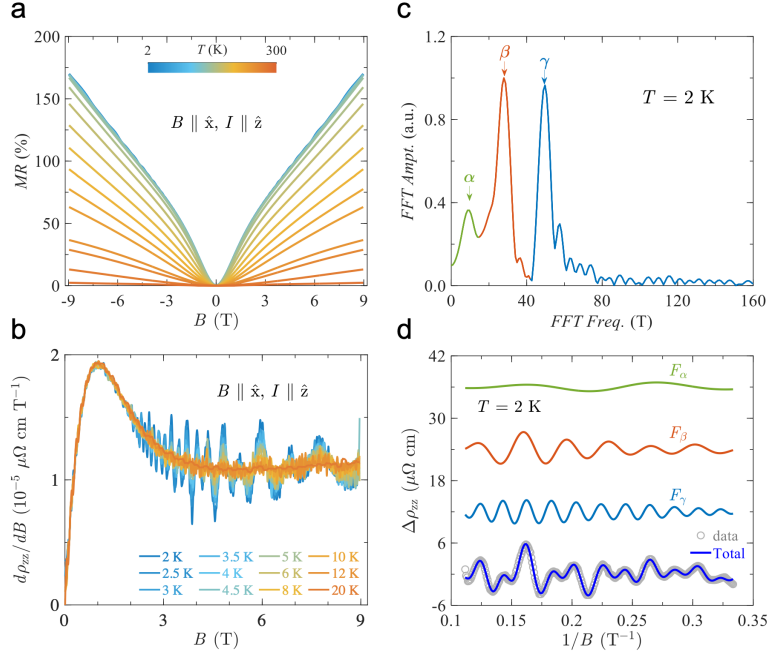

FIG. S6. Longitudinal Magnetoresistance (MR) and resistivity. (a) MR for different temperatures for $I \parallel \hat{z}$, $B \parallel \hat{x}$. (b) first derivative from magnetic field for $\rho_{zz}$ below 20 K for clarify the SdH oscillations. (c) Fast Fourier transformation (FFT) results of the observed SdH quantum oscillations of $\Delta\rho_{zz}$. Frequency range for reconstructing each oscillation are plotted in green ($\alpha$), orange ($\beta$), and azure ($\gamma$) colors. (d) Reconstructed SdH oscillations for $\alpha$-, $\beta$- and $\gamma$-pocket by bandwidth pass processes at 2 K.

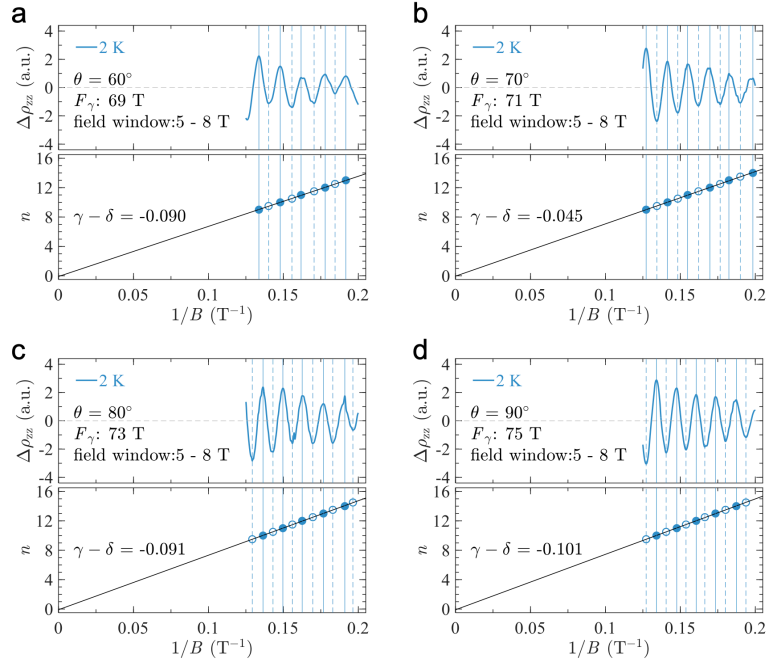

FIG. S7. SdH oscillations and Landau level (LL) fan diagram of $\gamma$-pocket at 2 K for various rotating angles ($B$ rotates from $\hat{x}$ to $\hat{z}$). (a-d) are the data of $\theta = 60°, 70°, 80°, 90°$, respectively. The LLs are assigned directly from the observed oscillations without filtering.

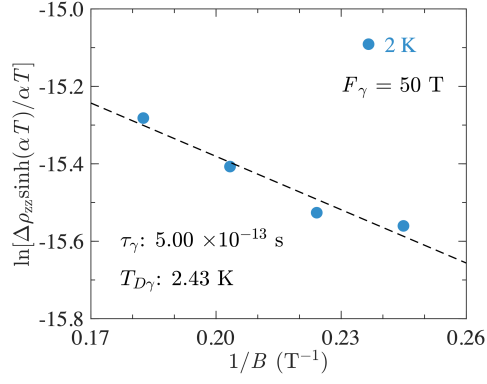

FIG. S8 Dingle plot of $F_\gamma$ when $B \parallel \hat{x}$.

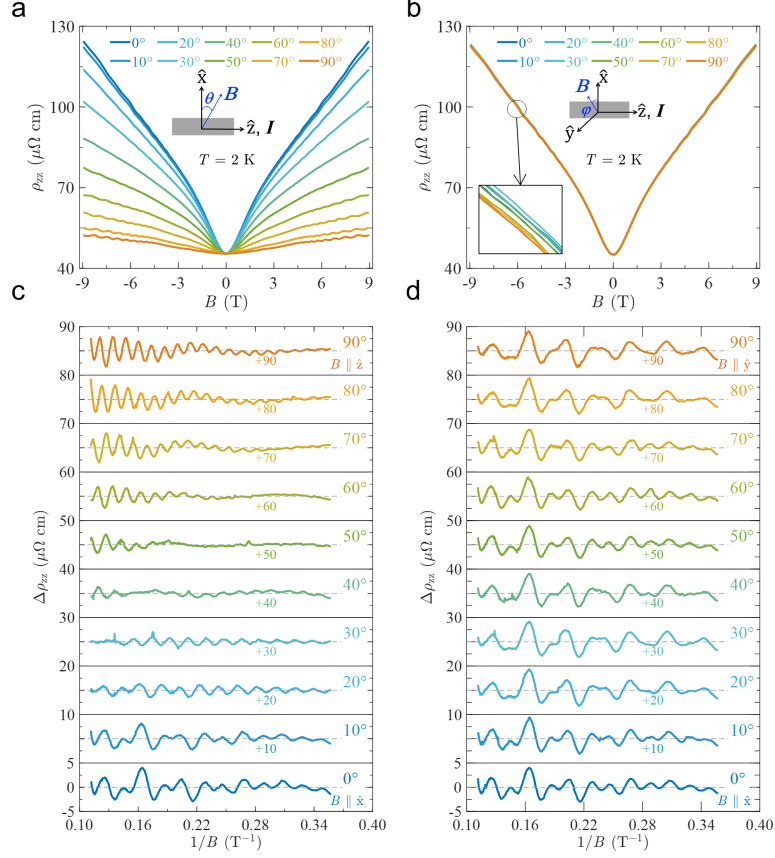

FIG. S9. Resistivity, SdH oscillations. (a) and (b) are resistivity when magnetic field is rotating from $\hat{x}$ to $\hat{z}$ and $\hat{x}$ to $\hat{y}$, respectively. The inset of (b) is the enlarged view for clarity. (c) and d, background subtracted resistivity $\Delta\rho_{zz}$ as functions of $1/B$ for the both configurations mentioned in (a) and (b) respectively, showing the periodic oscillations.

**Electronic band structure calculations**

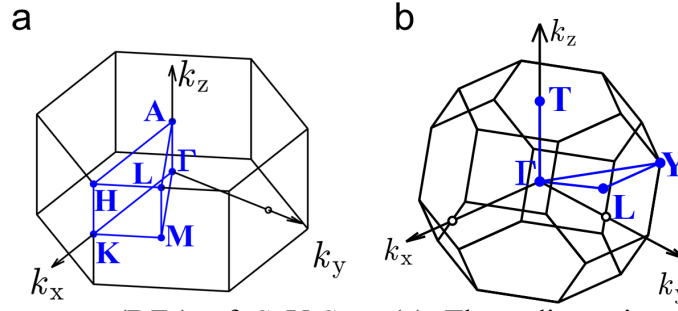

FIG. S10 Brillouin zones (BZs) of ScV$_3$Sn$_6$. (a) Three-dimensional (3D) BZ of high-temperature (280 K) phase of ScV$_6$Sn$_6$ with space group *P*6/*mmm*. **(b)** 3D BZ of low-temperature (50 K) phase of ScV$_6$Sn$_6$ with space group *R*32. Crystallographic information used in calculation are taken from the previous report [1].

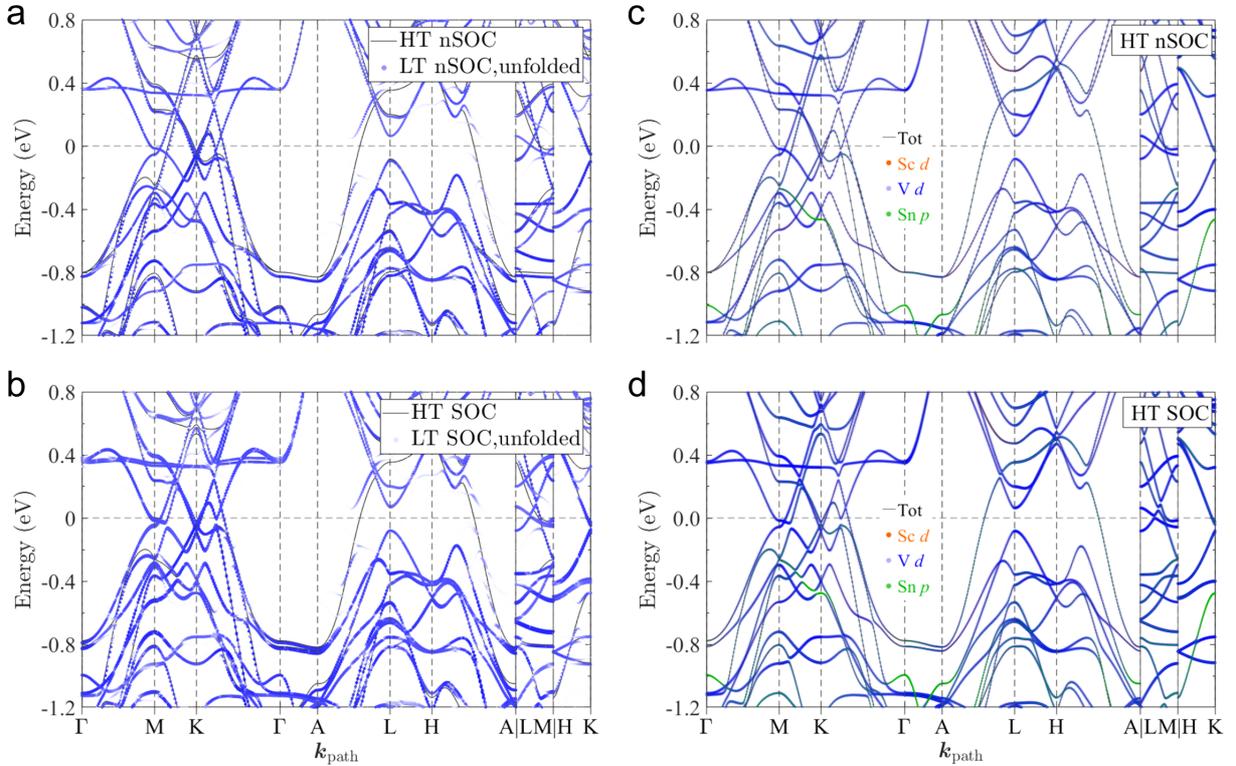

FIG. S11. Electronic band structures. (c) Band structure of ScV$_6$Sn$_6$ based on the crystal structure in high-temperature (HT) phase (black) and low-temperature (LT) CDW phase (blue) when spin-orbit coupling (SOC) is ignored. **(b)** Band structures with SOC. In a and b, the band structures of LT phase are unfolded to match the band structure of HT phase for comparison. (c, d) Orbitals projected band structure for HT phase with and without SOC, respectively.

We have calculated the band structures of ScV$_6$Sn$_6$ in the high temperature (HT) phase as well as in the low temperature (LT) CDW phase. The LT phase band structure is unfolded to match the one of HT phase when considering spin orbital coupling (SOC) (Fig. S11a) and neglecting SOC (Fig. S11b), respectively. The band structures didn't change a lot with or without considering SOC. Specifically, for LT phase, energy bands along *A-H-L-A* (i.e. $k_z = \pi$ plane) are gapped by CDW phase transition, which is consist with the results of angle-resolved photoemission spectrum (ARPES) [13]. The orbital projected band structure is also calculated and plotted in Fig. S11c and d. It is clear that the kagome V $d$ orbitals dominate the bands close to the Fermi level. Only a few bands along $M$-$K$, $G$-$A$, and $H$-$K$ are contributed by Sn $p$ orbitals, while the $d$ orbitals from Sc are negligible.

Table S1. Fermi surface parameters (effective mass $m^*$, Fermi vector $k_F$, Fermi velocity $v_F$, scattering length $l$, Dingle temperature $T_D$, quantum scattering time $\tau_q$, quantum mobility $\mu_q$) extracted from the SdH quantum oscillations from Figs. 3 and S6

| $F$ (T) | $m^*$ ($m_0$) | $A_F$ (Å$^{-2}$) | $k_F$ (Å$^{-1}$) | $v_F$ ($10^5$ ms$^{-1}$) | $l$ (nm) | $T_D$ (K) | $\tau_c$ ($10^{-13}$ s) | $\mu_c$ ($10^3$ cm$^2$V$^{-1}$s$^{-1}$) |
|---|---|---|---|---|---|---|---|---|
| $F_\alpha$ : 10 | 0.105 | 0.001 | 0.017 | 1.92 | - | - | - | |
| $F_\beta$ : 28 | 0.128 | 0.0027 | 0.029 | 2.64 | - | - | - | - |
| $F_\gamma$ : 50 | 0.184 | 0.0048 | 0.039 | 2.45 | 123 | 2.43 | 5.00 | 4.78 |

**References:**

[1] H. W. S. Arachchige, W. R. Meier, M. Marshall, T. Matsuoka, R. Xue, M. A. McGuire, R. P. Hermann, H. Cao, and D. Mandrus, *Charge Density Wave in Kagome Lattice Intermetallic ScV$_6$Sn$_6$*, Phys. Rev. Lett. **129**, 216402 (2022).

[2] J. Rodríguez-Carvajal, *Recent Advances in Magnetic Structure Determination by Neutron Powder Diffraction*, Physica B: Condensed Matter **192**, 55 (1993).

[3] G. Pokharel, S. M. L. Teicher, B. R. Ortiz, P. M. Sarte, G. Wu, S. Peng, J. He, R. Seshadri, and S. D. Wilson, *Electronic Properties of the Topological Kagome Metals YV$_6$Sn$_6$ and GdV$_6$Sn$_6$*, Phys. Rev. B **104**, 235139 (2021).

[4] S. Blundell, *Magnetism in Condensed Matter*, Reprint (Oxford Univ. Press, Oxford, 2014).

[5] B. R. Ortiz, L. C. Gomes, J. R. Morey, M. Winiarski, M. Bordelon, J. S. Mangum, I. W. H. Oswald, J. A. Rodriguez-Rivera, J. R. Neilson, S. D. Wilson, et al., *New Kagome Prototype Materials: Discovery of KV$_3$Sb$_5$, RbV$_3$Sb$_5$, and CsV$_3$Sb$_5$*, Phys. Rev. Materials **3**, 094407 (2019).

[6] N. Bansal, Y. S. Kim, M. Brahlek, E. Edrey, and S. Oh, *Thickness-Independent Transport Channels in Topological Insulator Bi$_2$Se$_3$ Thin Films*, Phys. Rev. Lett. **109**, 116804 (2012).

[7] F. C. Chen, Y. Fei, S. J. Li, Q. Wang, X. Luo, J. Yan, W. J. Lu, P. Tong, W. H. Song, X. B. Zhu, et al., *Temperature-Induced Lifshitz Transition and Possible Excitonic Instability in ZrSiSe*, Phys. Rev. Lett. **124**, 236601 (2020).

[8] J. Xu, Y. Wang, S. E. Pate, Y. Zhu, Z. Mao, X. Zhang, X. Zhou, U. Welp, W.-K. Kwok, D. Y. Chung, et al., *Unreliability of Two-Band Model Analysis of Magnetoresistivities in Unveiling Temperature-Driven Lifshitz Transition*, Phys. Rev. B **107**, 035104 (2023).

[9] Y. L. Wang, L. R. Thoutam, Z. L. Xiao, J. Hu, S. Das, Z. Q. Mao, J. Wei, R. Divan, A. Luican-Mayer, G. W. Crabtree, et al., *Origin of the Turn-on Temperature Behavior in WTe$_2$*, Phys. Rev. B **92**, 180402 (2015).

[10] M. N. Ali, L. M. Schoop, C. Garg, J. M. Lippmann, E. Lara, B. Lotsch, and S. S. P. Parkin, *Butterfly Magnetoresistance, Quasi-2D Dirac Fermi Surface and Topological Phase Transition in ZrSiS*, Sci. Adv. **2**, e1601742 (2016).

[11] X. Xu, X. Wang, T. A. Cochran, D. S. Sanchez, G. Chang, I. Belopolski, G. Wang, Y. Liu, H.-J. Tien, X. Gui, et al., *Crystal Growth and Quantum Oscillations in the Topological Chiral Semimetal CoSi*, Phys. Rev. B **100**, 045104 (2019).

[12] N. Huber, K. Alpin, G. L. Causer, L. Worch, A. Bauer, G. Benka, M. M. Hirschmann, A. P. Schnyder, C. Pfleiderer, and M. A. Wilde, *Network of Topological Nodal Planes, Multifold Degeneracies, and Weyl Points in CoSi*, Phys. Rev. Lett. **129**, 026401 (2022).

[13] S. Lee, C. Won, J. Kim, J. Yoo, S. Park, J. Denlinger, C. Jozwiak, A. Bostwick, E. Rotenberg, R. Comin, et al., *Nature of Charge Density Wave in Kagome Metal ScV6Sn6*, arXiv:2304.11820v1 (2023).